# SCIENTIFIC REPORTS

**OPEN**  Anomalous Tunnel Magnetoresistance and Spin Transfer Torque in Magnetic Tunnel Junctions with Embedded Nanoparticles

Arthur Useinov[1,2,3], Lin-Xiu Ye[2], Niazbeck Useinov[3], Te-Ho Wu[4] & Chih-Huang Lai[2]

The tunnel magnetoresistance (TMR) in the magnetic tunnel junction (MTJ) with embedded nanoparticles (NPs) was calculated in range of the quantum-ballistic model. The simulation was performed for electron tunneling through the insulating layer with embedded magnetic and non-magnetic NPs within the approach of the double barrier subsystem connected in parallel to the single barrier one. This model can be applied for both MTJs with in-plane magnetization and perpendicular one. We also calculated the in-plane component of the spin transfer torque (STT) versus the applied voltage in MTJs with magnetic NPs and determined that its value can be much larger than in single barrier system (SBS) for the same tunneling thickness. The reported simulation reproduces experimental data of the TMR suppression and peak-like TMR anomalies at low voltages available in leterature.

Various multilayer systems such as magnetic tunnel junctions (MTJs) and their modifications attracts considerable attention due to their promising applications. The current progress in studying the spin transfer torque, spin Hall effect, and TMR effect will provide the successful solutions of the problems related to the low critical switching current and the high thermal stability factor of magnetoresistive random access memory (MRAM) cells and other nanodevices[1].

It is well known that the double barrier system (DBS) might have a few advantages in relation to the single barrier system (SBS). As example, it was previously demonstrated that the double barrier MTJ is more suitable and applicable for MRAM due to larger $\Delta R/R$ ratio ($\Delta R$ is resistance change of the tunneling cell between two magnetic states (or bits), $R$ is a resistance of the junction)[2], as well as grater voltage output ($\Delta V = V \times$ TMR, where $V$ is applied voltage)[3,4]. Furthermore, it was reported that DBSs are riched with a quantum effects, e.g. oscillations of the tunneling conductance at low voltages which confirm an existance of the quantum well (QW) states in the middle layer[5]. The high exchange energy splittings of the conduction bands in Fe, Co, Ni and FeCoB, FeNi alloys that are used for MTJ fabrication provide strong differences between barrier transparencies for the majority and minority electrons. Considering that the electrons conserves their spin, the tunneling for the spin-up and spin-down electrons can be treated within the model of the two conduction channels[6,7], where electrons originating from a particular spin state of the one electrode tunnel into the empty states of the other one. Moreover, if the total thickness of the tunneling system is around 5 nm the electron *cannot* be considered as a particle. The Coulomb blockade effect is limited to the cotunneling and consecutive tunneling models[8–14], therefore, we assume that Coulomb blockade can be neglected. We considered the quantum solution for DBS at low temperatures and the generalized quantum-ballistic transport model which is similar to coherent tunnel model[15].

[1]Department of Physics, National Tsing Hua University, Hsinchu, Taiwan. [2]Department of Materials Science and Engineering, National Tsing Hua University, Hsinchu, Taiwan. [3]Institute of Physics, Kazan Federal University, Kazan, Russian Federation. [4]Graduate School of Materials Science, National Yunlin University of Science and Technology, Douliou, Taiwan. Correspondence and requests for materials should be addressed to C.-H.L. (email: chlai@mx.nthu.edu.tw)

In the present study, the transport problem was explored in terms of electron propagating waves reflecting the superposition of the quantum states for all magnetic layers, barriers and NPs within the double barrier approach[16] and generalized point-like contact model (see Supplementary Information). It is important to notice that the consecutive model and related Coulomb blockade due to basic definition cannot achieve the limit of the ballistic model. In fact, when we reduce the barrier thickness to the case that the NPs touch the both leads, the ballistic conduction occurs, which can only be explained by the point-contact model, instead of Coulomb blockade. We consider the system with small NP size and barrier thickness (in relation to mean free path of electron, $l \geq 15$ nm in metals) that Coulomb blockade cannot be achieved yet because electron has a wave quantum properties up to 5 nm–6 nm in nanoscale and our electron transport regime is before regime of the percolation. Percolation means consequent tunneling regime of the conductance with possible realization Coulomb blockade when consequent tunneling is suppressed (the electron shows more classical properties at this case).

Recent experimental and theoretical studies confirmed that double barrier magnetic tunnel junctions are promising structures for nanodevices[5,17–21], for instance Gao et al.[19] proposed the fabrication of a new kind of the spintronic devices such as the memory cell or STT-MRAM on the basis of the tunnel junctions, where the insulating layer is deposited as discontinuous media with embedded NPs. One of the simple ways to fabricate NPs in the plane of the insulating layer is plasma-assisted deposition with the low deposition rate. Nanoparticles grow up from the grains as a result of the atomic clusterization process[8,11,22].

Up to the present time there is an absence of systematically sufficient theoretical studies in the related field that explain the anomalous TMR behavior appropriately in the system with embedded NPs. The goal of this work is to present a simple approach explaining mainly all kinds of experimental observations, specifically those published by researchers from the IBM Almaden Research Center[11] and Taiwan SPIN Research Center[22]. These reports[11,22] are the main source of the experimental data for the comparison with our numerical simulations. In addition, our approach can be used to predict spin transfer torque in MTJs with embedded NPs.

In ref. 11, the MTJ with the in-plane magnetic anisotropy was constructed on the basis of $SiO_2/Ta(10)/Ir_{22}Mn_{78}(25)/Co_{70}Fe_{30}(3.5)/Mg(0.8)/MgO(2.8 nm)/Co_{70}Fe_{30}(7)/Ta(10)$ structure and the one which contains embedded $Co_{70}Fe_{30}$ NPs inside the MgO barrier: $SiO_2/Ta(10)/Ir_{22}Mn_{78}(25)/Co_{70}Fe_{30}(3.5)/Mg(0.8)/MgO(2.5)/Co_{70}Fe_{30}(t_{NP})/Mg(0.8)/MgO(2.5)/Co_{70}Fe_{30}(7)/Ir_{22}Mn_{78}(15)/Ta(10)$ (thicknesses in nm). These structures were characterized by the middle layer thickness $t_{NP}$, where $t_{NP}$ values from 0.25 to 0.75 nm which corresponds to the average NP (or nanodot) diameters from $1.53 \pm 0.4$ to $3.2 \pm 0.7$ nm, respectively. In fact, the NPs have the size dispersion, and its formation can be controlled by deposition rates, substrate materials and annealing.

The experimental measurements in refs 8,11 show the unusual TMR enhancement and the TMR suppression at the zero bias-voltage at low temperatures of a few kelvins. In particular, the TMR suppression has been explained by the Kondo-assisted tunneling regime with $t_{NP} < 0.7$ nm, while the low-bias peak-like TMR enhancement was observed for $t_{NP} \sim 1.1$ nm (average size of the NPs is larger). Results of the ref. 11 was treated as co-tunneling regime[23] with the assumption that the TMR anomalies were related to the Kondo resonant-tunneling. Furthermore the interpretation of these behaviors was given within the existence of sequential tunneling with Coulomb blockade, co-tunneling and non-resonant (and resonant) Kondo-assisted tunneling regimes, which mostly depend on the $t_{NP}$ value[11]. In addition, it was noted that the crossover from Kondo to co-tunneling behaviors for $t_{NP} \leq 1$ nm is correlated with the suppression threshold of the NP magnetic moments. If this moment is exist in NPs, then it could be coupled with one of the magnetic layers or vs. the external magnetic field. As a result, NPs have a temperature threshold (blocking temperature) when magnetic moment direction might be stable for finite low temperature while for higher temperature not stable. One of the experimental methods such as the field-cooled magnetization technique was used to indicate the dispersion of the NP blocking temperatures and observe the considerable fluctuations of the magnetic moments of the NPs persisting at low temperatures.

In the experimental work of Ye et al.[22] the MTJs with the perpendicular magnetic anisotropy (pMTJs) were constructed on the basis of $SiO_2/Ta(25)/Co_{40}Fe_{40}B_{20}(1.3)/MgO(1)/Fe(t_{NP})/MgO(1)/Co_{20}Fe_{60}B_{20}(2.2)/Ta(5)$ structures, which were measured for the fixed current values $I = 20\mu A$. As a result, the anomalous Hall voltages and $R$–$H$ loops were measured for various $t_{NP}$, since the Fe layer thickness $t_{NP}$ strongly influences the coercivity of the samples. Furthermore, the decrease in the coercivity with the further increase in the Fe content (varying $t_{NP}$ from 0.15 to 0.3 nm) is attributed to the percolation threshold and the formation of the connected network of magnetic granules as was performed previously[24]. Moreover, in comparison with the reference sample without Fe nanoparticles, one order enhancement of TMR value was discovered in the system with small $t_{NP} = 0.15$ nm. This behavior may originate from the possible enhancement of the spin polarization or spin current filtering on NP[25], but *not* due to Coulomb blockade effects.

The TMR behaviors, which were manifested in experiments[8,11,22], can be explained within presented quantum mechanical model, where only one coherent tunneling regime was considered. The TMR effect arises due to two different resistance states of the junction with parallel (P) and antiparallel (AP) magnetization alignments of the top and bottom magnetic layers, i.e. $TMR = (R^{AP} - R^P)/R^P$. The NPs may have its own independent magnetization directions, which influence the total resistance state. In our calculations, we consider two cases of the NP magnetic moment direction: co-aligned and anti-aligned with the magnetization direction of the free magnetic layer. The key parameter which describes the NP electron states is the spin-resolved wavenumber $k_{n\uparrow,\downarrow}$ (the wavenumber $k_n$ is not spin-resolved in the case of non-magnetic NPs). Some theoretical models[10,26] and experiments[27] show that NPs which are not magnetic at room temperature might have a weak magnetic properties and spin-resolved density of states ($\sim k_{n\uparrow,\downarrow}$) due to exchange interaction with magnetic layers at low temperature.

## Model of coherent tunneling

The electron transport model through the NP is similar to that for the double barrier magnetic tunnel junction (DMTJ). This model has a long history of applications[16,28–32] and development[33–36]. This theoretical works described the electron transport through a nanojunction between two different ferromagnetic leads taking into account the spin-dependent momentum conservation law. This theory utilizes quasi-classical and quantum mechanical approaches and is based on the solution of an extended kinetic equations for Green functions. Therefore it can be applied for the electron transport calculations through the tunnel barrier within a quantum-ballistic limit for single and double barrier MTJs[16,30,31].

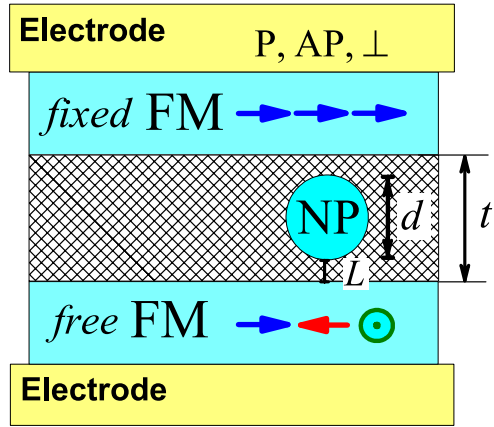

**Figure 1. Schematic view of the symmetric tunneling cell (TC) with NP.** The colored arrows show the parallel (P) $\gamma = 0$, anti-parallel (AP) $\gamma = \pi$ and perpendicular $\gamma = \pi/2$ magnetization of the ferromagnetic layers (FMs); $\gamma$ is angle between FMs magnetizations.

The working formula Eq. 1 for the conduction spin channel is coming from generalized point-contact model (the validation of the Eq. 1 can be found in Supplementary Information):

$$G_s^{P(AP)}(V) = G_0 \frac{\sigma \cdot k_{F,s}^2}{2\pi} \langle \cos(\theta_s) D_s^{P(AP)}(V) \rangle_{\theta_s, \phi} \quad (1)$$

The conductance $G = I/V$ is proportional to the product of the transmission $D_s^{P(AP)}$ and cosine of the incidence angle of the electron trajectory $\cos(\theta_s)$ with the spin index $s = (\uparrow, \downarrow)$; $k_{Fs}$ is the wavenumber on the Fermi level of the top (bottom) FM layers as a function of the positive (negative) sign of the applied voltage $V$; for simplicity, the top and bottom FM layers are taken as equivalent to each other (see Fig. 1), except magnetic properties (bottom layer is magnetically soft, see Fig. 1); $G_0 = e^2/h$ is the spin-resolved conductance quantum $G_0 \approx 3.874 \times 10^{-5} \Omega^{-1}$ and $\sigma$ is cross section area factor. Angular brackets $\langle \ldots \rangle_{\theta_s, \phi}$ in Eq. 1 are determined as follows: $\langle \ldots \rangle_{\theta_s, \phi} \equiv \frac{1}{2\pi} \int_0^{\theta_{\min}} \sin(\theta_s) d\theta_s \int_0^{2\pi} d\phi(\ldots) = \int_{X_{CR,s}}^{1.0} x_s D_s^{P(AP)} dx_s$, where $x_s = \cos(\theta_s)$. The lower limit $X_{CR,s} = \sqrt{|1 - [k_{F,s'}(V)/k_{F,s}]^2|}$ gives the critical angle restriction $\theta_{\min} = \arccos(X_{CR})$ when the electron tunnels from the conduction band of the top FM layer with $k_{Fs}$ into the minority or majority bands of the bottom layer with $k_{Fs'}(V)$. Depending on the properties of the materials the majority band can be characterized by $k_{F\uparrow}$ or $k_{F\downarrow}$, but in most FMs it corresponds to $k_{F\uparrow}$.

In DBS simulation we used analytical solution for the transmission, in this case point-contact area exchanged by the system, which has double barrier energy potential profile and cross section area of NP $\sigma = \sigma_0 = \pi(d/2)^2$. The transmissions $D_s^{P(AP)}$ of the conduction electrons through the DBS were calculated according to the quantum mechanical regulations. The solution is stationary over time and space. Analytical views of the transmission are accessed in refs 16,37. SBS was derived as a limit of the DBS at $d \to 0$. The cross section area factor for SBS was determined as $\sigma = (Q - \sigma_0)$, where $Q$ is a unit area of the tunneling cell.

Here we employed the parallel circuit connection of the unit tunneling cells, where each cell contained one size-averaged NP per unit cell area $Q$, which characterizes concentration of the NPs in MTJ. The tunnel junction with the surface area $S$ consists of $N = S/Q$ cells. For the real systems, however, it is important to consider the size distribution of NPs and calculate the conductance according to this distribution. Therefore only the size-averaged NP was considered as a basic model here.

The total conductance of the junction is $G = N \times \Sigma_s(G_{1,s} + G_{2,s})$, where $G_{1,s}$ is the spin-up and spin-down conductance through NP (coherent double barrier tunneling with equal barriers). $G_{1,s}$ is the dominant term and $G_{2,s}$ is the conductance of the direct tunneling through the insulator without NP. Finally, TMR = $(G^P - G^{AP})/G^{AP} \times 100\%$.

It is necessary to notice that consecutive tunneling regime is not considered since its impact is much less than $G_1 = G_{1,\uparrow} + G_{1,\downarrow}$. The consecutive regime might be only dominant in MTJs with embedded NPs for $d > 3.5$ nm ($t_{NP} > 1.5$ nm), and for granular systems above the percolation threshold, e.g. in NPs/I multilayers [Fe(0.1 < $t_{Fe}$ < 1.5 nm)/MgO($t_{MgO}$ nm)]$_N$ according to ref. 24.

The present study shows that the formalism, which was developed for DMTJs is applicable for the model of tunnel junctions with embedded magnetic and non-magnetic NPs. The conducting energy states of metallic NPs at low temperatures can be quantized and approximately satisfy the QW solution for the energy $E_n = \hbar^2 k_n^2/2m$, where the electron wavenumber can be estimated as follows:

$$k_n = n\pi/d \quad (2)$$

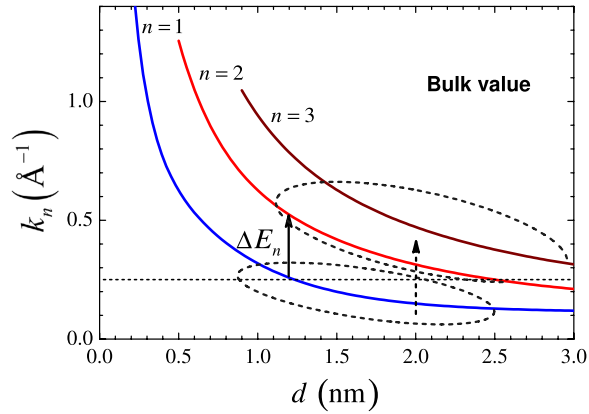

**Figure 2. Quantum well solution for the $k_n = n\pi/d$ with $n = 1, 2, 3$.** Horizontal dashed line shows an example of the fixed $k_n = 0.251$ Å$^{-1}$ corresponding to the MTJs with NP $d = 1.25$ nm, $n = 1$ or $d = 2.5$ nm, $n = 2$. Arrows show possible switching between QW states.

where $n = 1, 2, 3, \ldots$ (see Fig. 2); the NP diameter $d$ assumed to be equal to the width of the QW, the distance between barriers and proportional to the number of atoms inside NP. The set of $k_{F,s}$ and $k_{n,s}$ for FM layers and NPs are initial parameters, respectively which characterize the state of the system at $V \simeq 0$ ($V = 10^{-4}$V) and temperature $T \to 0$. In contrary the state with $V \neq 0$ modifies the absolute values of the wavenumbers according to the following relations: $k_{F,s'}(V) = \sqrt{k_{F,s'}^2 + c \cdot V}$, neglecting the voltage drop on the electrodes and FMs, and:

$$k_{n,\uparrow(\downarrow)}(V) = \sqrt{k_n^2 \pm m\Delta_0/\hbar^2 + cM \cdot V/2} \qquad (3)$$

$M = m/m_0$ is the ratio of the effective mass $m$ to the free electron mass $m_0$, $s' = \uparrow(\downarrow)$, $c = 2m_0 e/\hbar^2 \cdot 10^{-20}$ [Å$^{-2}$ eV$^{-1}$] is the dimensional factor, $\Delta_0$ is a small energy difference between the spin-up and spin-down electron states in the case of magnetic NPs.

The approach of the NP's energy levels in range of the QW solution (Shrödinger equation) is the most simplified picture and is used in the model to show an approximate scale for $k_n(d)$ and $n$, however we ignore discrete QW solution in some cases. More accurate energy bands (electronic spin band structure) and it's quantization (the gaps between states) can be found within different theoretical techniques such as R. Kubo formalism and random matrix theory[38,39], *abinitio* calculations or Landau perturbation theory[40]. For example *abinitio* calculation of the density of states was performed for TMR simulation in ref. 36.

## Results

**A. Anomalous TMR in MTJ with embedded NPs.** The anomalous TMR behaviors are shown in Fig. 3. It reproduces the peak-like TMR and its suppression at zero voltage (TMR$_0$), which was demonstrated previously[8,11,41]. TMR$_0$ can be even negative. According to our simulations, the solution was determined with assumption $T \to 0$. The electron energy is equal to $E_F$ and $E_F \gg k_B T$. The initial parameters in Fig. 3 were taken as follows: $k_{F\downarrow} = 0.421$ Å$^{-1}$, $k_{F\uparrow} = 1.09$ Å$^{-1}$, ($k_{F\uparrow} = 1.2$ Å$^{-1}$ for curves 3 and 5, Fig. 3a), $M = 0.8$, both barrier heights $U_B = 1.2$ eV, both barrier thickness $L_{1,2} = 1$ nm, effective mass for barriers $0.4m_0$ and $Q = 20.0$ nm$^2$.

The most strong TMR$_0$ suppression is related to $n = 1$ within the QW solution $k_n = n\pi/d$ (Fig. 2). In the case of the occupied states $n = 1$ ($d = 1.25$) or $n = 3$ ($d = 2.6$) the TMR behavior was switched to the classical one, which is similar to those in planar DMTJ[16,29]. Figure 2 shows the QW solution for the $k_n(d)$ with $n = 1, 2, 3$, where $\Delta_0 = 0$, $V = 0$. For example, solid black arrow shows the switching $n = 1$ into $n = 2$ state with the energy gap $\Delta E_n = \hbar^2/2m(k_{n=2}^2 - k_{n=1}^2) = 0.9$eV for $d = 1.25$ nm, and it corresponds to the transition between 1 and 4 TMR curves in Fig. 3a. The energy gap between all states rapidly becomes smaller when $d$ grows, while the Fermi level of the NP increases.

The suppressed TMR$_0$ behavior (corresponding to small wavenumber $k_n \approx 0.1 - 0.36$ Å$^{-1}$) was observed only at low temperatures in experiments (the temperature related details is discussed in Supplementary Information). Furthermore, the TMR$_0$ achieves negative value at $k_n < \pi/d$ (Fig. 3c, curve 2) that might be possible due to the NP's shape anisotropy, interface effects, non-equal barrier thicknesses and other reasons and resulting in significant deviation $\Phi$ from the ideal QW approach $k_n = n\pi/d - \Phi$.

The simulations for $D_s^{P\,(AP)}$ at small $k_n$ correspond to the rapid tunnel transparency growths due to opened and permitted section of the trajectory angle $\theta_s$ at the threshold voltage, since $k_n(V)$ increases with voltage (Eq. 3). Thus, $G^P$ and $G^{AP}$ form a step-like behavior, Fig. 3d. The phenomenon is denoted as the quantized conductance regime due to restricted NP geometry. Once the conductance steps are located directly in the region of low voltage $V \approx 10^{-4}$ V the TMR$_0$ suppression takes place here.

The second kind of TMR anomalies such as peak-like TMR behavior is shown for $d = 2.6$ nm at $k_n = 0.4508$ Å$^{-1}$, (Fig. 3c curve 1). As can be seen the peak width within a few mV is similar to experimental one (Fig. 3b curve 4*). The origin of the TMR peak is the beginning of the quantized conductance regime due to the restricted contact geometry (Fig. 3d). Well-defined peak was discovered in very narrow range $0.4504 < k_n < 0.4509$ Å$^{-1}$, that reflect resonant TMR character and also rapidly increased resistance $R^{P(AP)} = 1/G^{P(AP)}$. The minor resonant TMR peaks and dips can be also observed at finite voltages (Fig. 3b curve 3 and Fig. 4b, see $d = 2.4$, $d = 1.8$ nm).

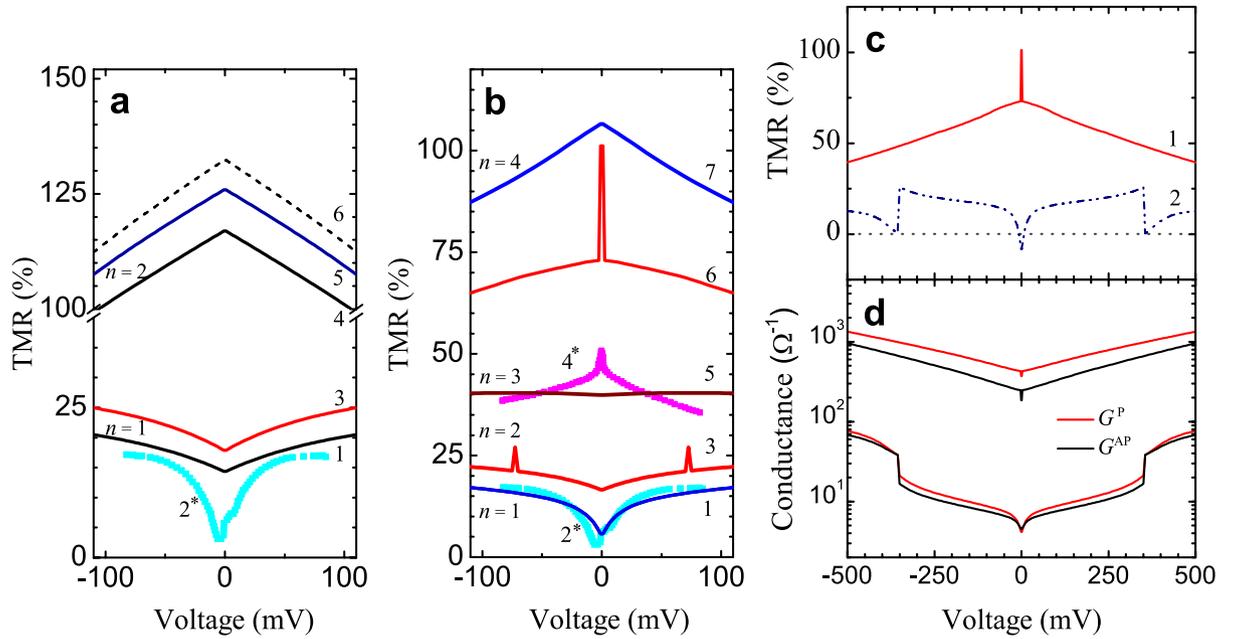

**Figure 3.** (**a**) presents TMR-$V$ simulations for $d = 1.25$ nm, $k_{n=1} = 0.251$ Å$^{-1}$ and $k_{n=2} = 0.5026$ Å$^{-1}$, curves 1 (3) and 4 (5) correspond to $k_{F,\uparrow} = 1.09$ (1.2) Å$^{-1}$, respectively. The dashed curve 6 correspond to the magnetic state of NP with $k_{F,\uparrow} = 1.09$ Å$^{-1}$, $k_{n=2,\uparrow} = 0.5051$ Å$^{-1}$, $k_{n=2,\downarrow} = 0.5001$ Å$^{-1}$, other parameters see in the text; Curves $2^*$ and $4^*$ in (**a**,**b**) is experimental data[11] for $t_{NP} = 0.45$ nm and $t_{NP} = 1.2$ nm, respectively ($T = 2.5$ K). (**b**) TMR curves 1, 3, 5–7 for the case of $d = 2.6$ nm correspond to the $k_n = 0.121$, 0.242, 0.362, 0.4508, 0.483 Å$^{-1}$, respectively. (**c**) Curve 1 shows peak-like TMR within extended voltage range and coincides with the curve 6 depicted in Fig. 3b, curve 2 shows negative TMR$_0 = -8\%$ for $k_n = 0.115$ Å$^{-1}$, $d = 2.6$ nm. (**d**) Corresponding step-like quantized conductance is shown.

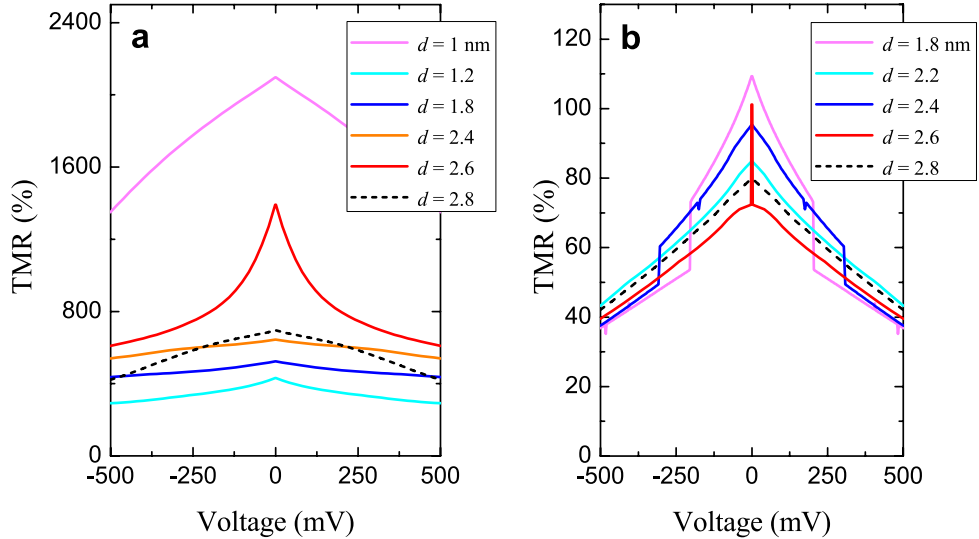

**Figure 4.** TMR curves according to the NP size distribution, where $k_{F,\uparrow} = 1.09$ Å$^{-1}$, $k_{F,\downarrow} = 0.421$ Å$^{-1}$. For all curves $k_n$ values are fixed: (**a**) $k_n = 1.1$ Å$^{-1}$; (**b**) $k_n = 0.4508$ Å$^{-1}$, other parameters are the same as in previous cases.

As a result the resonant TMR peak observed due to quantized behavior of the spin-polarized conduction channels for P and AP magnetic configurations: some of them are still closed for AP while partly open for P configuration, i.e. some section of the permitted $\theta_s$ is more restricted for AP against P state. The transmission probability amplitude is coherently suppressed for both P and AP states due to the imbalanced matching of the electron wave-functions on the interfaces. The resonant TMR$_0$ suppression means inversed situation, i.e. P state is more restricted against AP. Positive (negative) TMR value was obtained for $G^P > G^{AP}$ ($G^P < G^{AP}$).

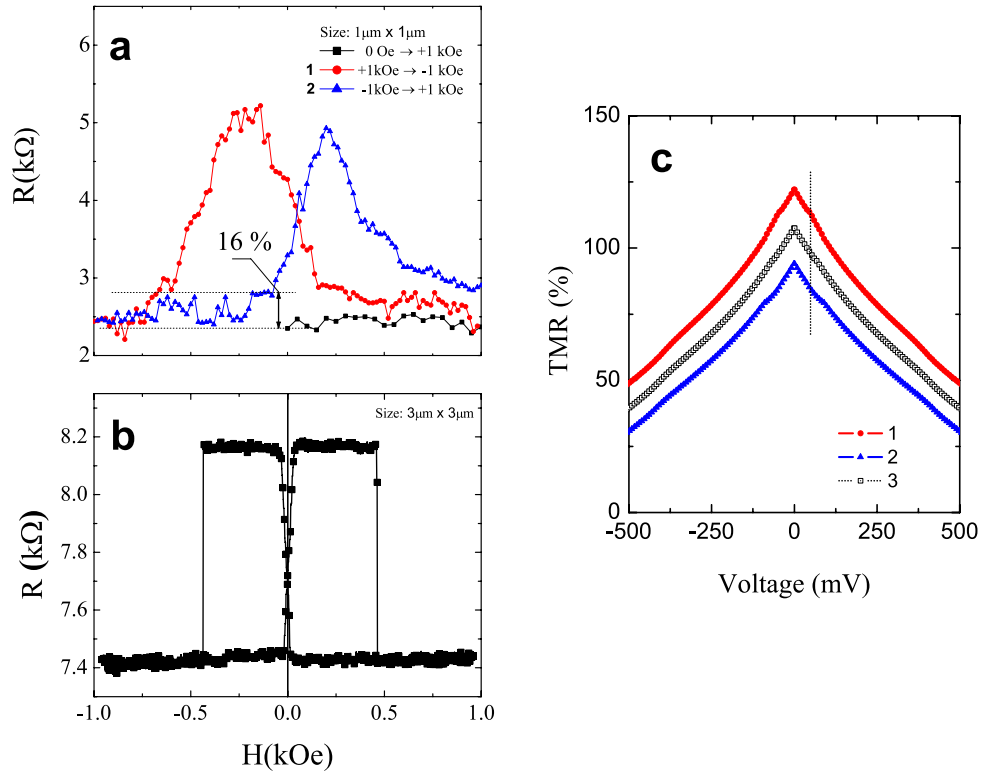

**Figure 5.** (**a**,**b**) show *R*–*H* loops of the experimental data with and without embedded NP at room temperature, respectively; (**c**) shows around 23% of the relative TMR change between curves 1 and 2. Initial parameters are as follows: $k_{F\uparrow} = 1.01\,\text{Å}^{-1}$, $k_{F\downarrow} = 0.45\,\text{Å}^{-1}$, $d = 1.82\,\text{nm}$, $k_{n,\uparrow(\downarrow)} = 0.54 \pm 0.002\,\text{Å}^{-1}$ for curve 1; $k_{n,\uparrow(\downarrow)} = 0.54 \mp 0.002\,\text{Å}^{-1}$ for curve 2, and curve 3 is *non*-magnetic case: $k_n = 0.54\,\text{Å}^{-1}$.

Interestingly, the observed quantized conductance and related resonant TMR are close to ballistic conductance[42] and giant magnetoresistance effect in nanocontacts[32,43], respectively. Therefore from this point of view, anomalous TMR is giant TMR effect.

The distribution of the TMR-*V* curves for different NP sizes is given in Fig. 4, where Fig. 4a correspond to $k_n = 1.1\,\text{Å}^{-1}$ (NP as a bulk limit) and Fig. 4b shows curves $k_n = 0.4508\,\text{Å}^{-1}$. For example, in Fig. 4a the $TMR_0$ demonstrates enhancement for $d = 1.0$ and $d = 2.6\,\text{nm}$. It can be noticed that the dominant impact on the peak-like TMR formation in real junctions with NP dispersion is probably coming from the NPs with $d > 1.5\,\text{nm}$ since the transmission channels based on NPs with $d < 1.5\,\text{nm}$ are much more resistive which make their contribution in total current less important, thus reflecting the experimental conditions $1.05 < t_{NP} < 1.35\,\text{nm}$ for the TMR peak observation[11]. In real tunnel junctions, the problem of the non-homogeneous voltage drop arises in the case of the asymmetric DBS with non-equal barriers[37].

**B. MTJ with magnetic NPs.** Since the anisotropy effects such as voltage-induced screening or density of states dependence of the NP and FM layers due to crystallographic order are not presented in this model, there is no difference between pMTJ and in-plane MTJ due to the fact that transmission does not depend on the choice of the quantization axis. All TMR effects including anomalous one occurs in pMTJs too.

In this section we analyzed fabricated pMTJs according to our experimental data[22]. In contrast to the maximal resistance switching related to P and AP magnetic configurations, the unstable minor resistance switching at room temperature was well-defined in the *R*–*H* loops as can be seen in Fig. 5a, the data taken from ref. 22. It is clearly observed back and forth switching of the minor resistance amplitudes by around 16% ($\simeq 0.4\,\text{k}\Omega$) with respect to total resistance change. It is assumed that this resistance switching is related to the reversal of the magnetization of almost all nanoparticles which co-oriented or anti-oriented with respect to the free layer magnetization direction. The bottom layer is a free (soft) layer in the samples, which is similar to presented model.

In the case of our data[22] (Fig. 5a), pMTJ with NPs has the field offset of the *R*–*H* loop with respect to that of pMTJ without NPs (Fig. 5b). Such conversion between AP and P states is allowed due to the stronger magneto-static interaction between FMs layers and magnetic NPs. It induces the large amount of intermediate magnetic configurations of the domains in the soft layer.

The curve 1 (red circles) shown in Fig. 5a for the NP $d = 1.82\,\text{nm}$ corresponds to the TMR with the amplitude of 111% at $I = 20\,\mu\text{A}$, while curve 2 (blue triangles) represent TMR $\approx 85\%$ that perfectly correlates with our theoretical estimations, Fig. 5c. For comparison, Fig. 5b shows the *R*–*H* loops of the pMTJ without NPs. The TMR ratio is around 11% at room temperature, and the sharp resistance switching is observed between P and AP configurations. The same TMR value at similar conditions was also derived in ref. 44 but in contrast to present work the magnetization of the soft FM layer changed gradually even though the field offset of the R–H curve was also observed.

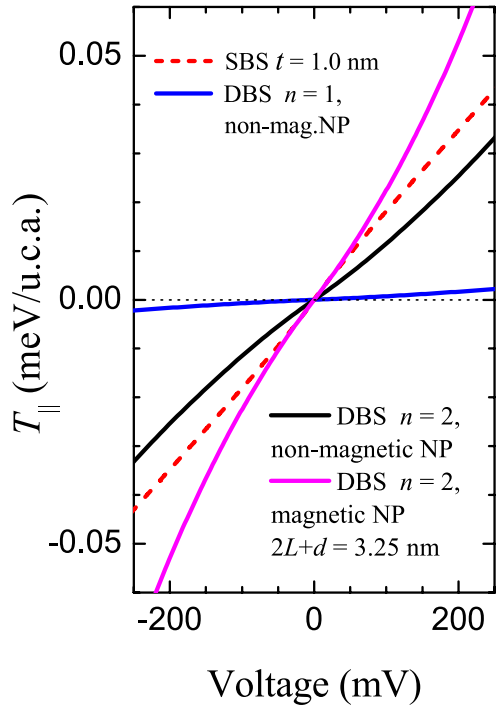

**Figure 6. Spin torque simulations for single and double barrier systems per unit cell area (u.c.a.) for different quantum states.** The parameters are similar to those for curves 3 and 5 in Fig. 3a, magenta curve is the STT-V dependence for the magnetic NP: $k_{n,\uparrow(\downarrow)} = 0.5026 \pm 0.025\,\text{Å}^{-1}$.

Figure 6c shows simulations of the magnetic NP for $d = 1.82$ nm. As can be seen, the $TMR_0$ increases to 122% (curve 1) and 92% (curve 2) at zero voltage, however, the vertical dashed line shows almost the same TMR values as experimental one at $V = 50$ mV. In this case we used following parameters: $k_{n,\uparrow} = 0.542\,\text{Å}^{-1}$, $k_{n,\downarrow} = 0.538\,\text{Å}^{-1}$ for curve 1, which is attributed to the co-aligned magnetic moment of the NP with respect to the free magnetization direction, and for curve 2 $k_{n,\uparrow} = 0.538\,\text{Å}^{-1}$, $k_{n,\downarrow} = 0.542\,\text{Å}^{-1}$ with the anti-aligned magnetic moment. The $k_n$ values are comparable with the $k$-values of the QW state above $n = 3$. Equation 3 was simplified to the $k_{n,s}(V) = \sqrt{(k_{n,s})^2 + cM \cdot V/2}$. The curve 3 shows the non-magnetic case $k_n = 0.54\,\text{Å}^{-1}$, other parameters such as effective masses and barrier's heights were taken from previous simulations.

As a result, the TMR amplitude depends on aligned or anti-aligned coupling of the NP magnetic moment. The gap in $0.004\,\text{Å}^{-1}$ between $k_{n,\uparrow}$ and $k_{n,\downarrow}$ is an approach of magnetic state of the NP with presence of the external magnetic field. Initial state of NP is assumed to be paramagnetic. Spin-resolved $k_n$ opens additional spin-resolved electron scattering on NP. Spin-up and spin-down electrons leads to increased variations of the tunnel transparencies. However in real systems there are always several reasons that reduce the theoretically estimated TMR values: screening effect, interface and internal disorder, spin-flip scattering, intermediate magnetic configurations of the domains, etc.

After simulations of the anomalous TMR and splitted TMR behavior in case of MTJ with magnetic NP, we used the same approach to calculate spin transfer torque because it is quite important spin dynamic characteristics for applications.

**C. Spin transfer torque.** One of the important effects which may characterize the tunnel junction is spin transfer torque[45–47]. Figure 6 shows estimation of the in-plane STT values (STT component along barrier interface) for DBS and SBS. The strong and almost linear enhancement of the in-plane torque vs. voltage is found for tunnel junctions with NPs ($n = 2$, $d = 1.25$ nm). It is shown that even if tunneling thickness for DBS is large, i.e. 3.25 nm ($t = 2L + d$), the STT is comparable to the value in SBS with $t = 1.0$ nm. The STT value calculated for the SBS with $t = 3.25$ ($d \to 0$ nm) is negligibly small, i.e. $T_\parallel \simeq 10^{-6}$ meV/Q at $V = 250$ mV (not shown). Moreover, STT is higher in case of magnetic NP (Fig. 6). Thus, we conclude that in the case of MTJs with magnetic NPs the larger STT effect should increase the fluctuations (blocking temperature) of the both magnetic moments for NPs and soft magnetic layer and it should also decrease the critical switching current value in STT-MRAM cell, some experimental evidence of this fact can be found in ref. 48.

The theoretical background, which assists to calculate the STT value by means of only charge current density components in the SBS was derived in refs 49,50:

$$T_{\parallel} = \frac{\sin(\gamma)}{2}(\mathfrak{J}^{P} - \mathfrak{J}^{AP}), \qquad (4)$$

where $\mathfrak{J}^{P(AP)} = \hbar(J_{\uparrow}^{P(AP)} - J_{\downarrow}^{P(AP)})/2e$ are the spin current densities. $J_{\uparrow\downarrow}^{P(AP)}$ are the charge current densities, which were derived as $J_s^{P(AP)} = (G_{1,s}^{P(AP)} + G_{2,s}^{P(AP)})V/Q$ and $G_{1(2),s}^{P(AP)}$ were calculated according to Eq. 1. $T_{\parallel}$ was calculated for the case when top and bottom FM magnetizations are perpendicular to each other ($\gamma = \pi/2$) and reach the maximal value, assuming magnetization direction of the NP is the same direction as the bottom FM layer. The STT was calculated in a similar way following Eq. 4 for single and double barrier systems due to the same definition of the electrical circuit in range of the coherent tunneling.

## Conclusion

In this work demonstrated theoretical technique allows us reproduce TMR anomalies according to the experimental studies available in literature. The model of the tunnel junctions with embedded NPs is based on the approach of the double barrier magnetic tunnel junction, which was developed earlier, however, in order to observe quantized conductance regime the role of the lowest QW states and small values of the wavenumber was crucially emphasized. The TMR anomalies such as the TMR suppression or peak-like TMR behavior at low voltages are simulated within coherent tunneling regime by tuning only one parameter ($k_n$ value) without consideration of the particle distribution by size. Moreover, the TMR amplitude variation was studied in pMTJs with magnetic NPs for two possible coupling cases such as co-aligned and anti-aligned with respect to the magnetization of the soft layer. Co-aligned coupling provided valuable TMR-$V$ amplitude enhancement. The present model predicted also large STT effect in tunnel junctions with magnetic embedded nanoparticles, which may induce the valuable reduction of the critical current density for current-induced magnetization switching.

The results show that considered junctions are complicated systems for simulations, where the implementation of the resonant TMR as well as the energy-effective STT magnetization switching can be achieved due to presence of nanoparticles. The developed theoretical technique is promising as a background, which makes it possible to generalize the model and describe the TMR behaviors more precisely with the NP dispersion. Moreover presented model can serve as a main guideline for the non-linear I-V curve calculations in novel spintronic and nanoscale quantum structures. In addition the approach can be applied for data analysis of the scanning probe microscope with conducting tip.

## References

1. Gallagher, W. J. & Parkin, S. S. P. Development of the magnetic tunnel junction MRAM at IBM: From first junctions to a 16-Mb MRAM demonstrator chip. *IBM J. Res. Dev*. **50,** 5–23 (2006).
2. Inomata, K., Saito, Y., Nakajima, K. & Sagoi, M. Double tunnel junctions for magnetic random acces memory deviced. *J. Appl. Phys.* **87,** 6064 (2000).
3. Gan, H. D. *et al*. Tunnel magnetoresistance properties and film structures of double MgO barrier magnetic tunnel junctions. *Appl. Phys. Lett.* **96,** 192507 (2010).
4. Feng, G. *et al*. Annealing of CoFeB/MgO based single and double barrier magnetic tunnel junctions: tunnel magnetoresistance, bias dependence, and output voltage. *J. Appl. Phys.* **105,** 033916 (2009).
5. Nozaki, T., Tezuka, N. & Inomata, K. Quantum oscillation of the tunneling conductance in fully epitaxial double barrier magnetic tunnel junctions. *Phys. Rev. Lett.* **96,** 027208 (2006).
6. Campbell, I., Fert, A. & Pomeroy, A. Evidence for two current condition in Iron. *Philos. Mag.* **15,** 977 (1967).
7. Mathon, J. & Umerski, A. Theory of tunneling magnetoresistance in a disordered FeMgOFe(001) junction. *Phys. Rev. B* **74,** 140404(R) (2006).
8. Ciudad, D. *et al*. Competition between cotunneling, Kondo effect, and direct tunneling in discontinuous high-anisotropy magnetic tunnel junctions. *Phys. Rev. B* **85,** 214408 (2012).
9. Feng, G., Dijken, V. S. & Coey, J. M. D. MgO-based double barrier magnetic tunnel junctions with thin free layers. *J. Appl. Phys.* **105,** 07C926 (2009).
10. Pustilnik, M. & Glazman, L. I. Kondo effect in quantum dots. *Phys.: Condens. Matter* **16,** R513–R537 (2004).
11. Yang, H., Yang, S. H. & Parkin, S. S. P. Crossover from Kondo-assisted suppression to co-tunneling enhancement of tunneling magnetoresistance via ferromagnetic nanodots in MgO tunnel barriers. *Nano Lett.* **8,** 340–344 (2008).
12. Nah, S. & Pustilnik, M. Kondo temperature of a quantum dot. *Phys. Rev. B* **85,** 235311 (2012).
13. Beenakker, C. W. J. Theory of Coulomb-blockade oscillations in the conductance of a quantum dot. *Phys. Rev. B* **44,** 1646 (1991).
14. Averin, D. V., Korotkov, A. N. & Likharev, K. K. Theory of single-electron charging of quantum wells and dots. *Phys. Rev. B* **44,** 6199 (1991).
15. Wilczynski, M. & Barnas, J. Tunnel magnetoresistance in ferromagnetic double-barrier planar junctions: coherent tunneling regime. *J. Mag. Magn. Mat.* **221,** 373–381 (2000).
16. Useinov, A., Kosel, J., Useinov, N. Kh. & Tagirov, L. R. Resonant tunnel magnetoresistance in double-barrier planar magnetic tunnel junctions. *Phys. Rev. B* **84,** 085424 (2011).
17. Ikegawa, S. *et al*. A fully integrated 1 kb magnetoresistive random access memory with a double magnetic tunnel junction. *Jpn. J. Appl. Phys.* **42,** L745–L747 (2003).
18. Iovan, A. *et al*. Spin diode based on Fe/MgO double tunnel junction. *Nano Lett.* **8,** 805–809 (2008).
19. Gao, C. H., Yang, Y. X., Xiong, Y. Q. & Chen, P. Low critical current density for spin-transfer torque in Fe-MgO granular film at room temperature. *J. Phys. D: Appl. Phys.* **47,** 045003 (2014).
20. Krishnamurthy, S., Chen, A. & Sher, A. Alignment of two-valley resonance levels and IV characteristics of GaAs/AlAs resonant tunneling diodes. *J. Appl. Phys.* **84,** 5354 (1998).
21. Chshiev, M., Stoeffler, D., Vedyayev, A. & Ounadjela, K. Magnetic diode effect in double-barrier tunnel junctions. *Europhys. Lett.* **58,** 257–263 (2002).
22. Ye, L.-X. *et al*. Embedded Fe nanoparticles in the MgO layer of CoFeB/MgO/CoFeB perpendicular magnetic tunnel junctions. *IEEE Trans. on Magn.* **50,** 4401203 (2014).
23. Takahashi, S. & Maekawa, S. Effect of coulomb blockade on Magnetoresistance in ferromagnetic tunnel junctions. *Phys. Rev. Lett.* **80,** 1758–1761 (1998).
24. García-García, A. *et al*. Tunneling magnetoresistance in Fe/MgO granular multilayers. *Apl. Phys. Lett.* **107,** 033704 (2010).

25. Pham, T. V. *et al.* Spin-dependent tunneling in magnetic tunnel junctions with Fe nanoparticles embedded in an MgO matrix. *Solid State Commun.* **183,** 18–21 (2014).
26. Martinek, J. *et al.* Kondo Effect in the Presence of itinerant-electron ferromagnetism studied with the numerical renormalization group method. *Phys. Rev. Lett.* **91,** 247202 (2003).
27. Koide, T. *et al.* Direct determination of interfacial magnetic moments with a magnetic phase transition in Co nanoclusters on Au(111). *Phys. Rev. Lett.* **87,** 257201 (2001).
28. Deminov, R. G., Useinov, N. Kh. & Tagirov, L. R. Magnetic and superconducting heterostructures in spintronics. *Magnetic Resonance in Solids. Electronic Journal* **16,** 14209 (2014).
29. Useinov, A. & Kosel, J. Spin asymmetry calculations of the TMR-V curves in single and double-barrier magnetic tunnel junctions. *IEEE Trans. on Magn.* **47,** 2724–2727 (2011).
30. Useinov, A., Mryasov, O. & Kosel, J. Output voltage calculations in double barrier magnetic tunnel junctions with asymmetric voltage behavior. *J. Mag. Mag. Mat.* **324,** 2844–2848 (2012).
31. Useinov, A. N., Deminov, R. G., Useinov, N. Kh. & Tagirov, L. R. Tunneling magnetoresistance in ferromagnetic planar hetero-nanojunctions. *Phys. Status Solidi B* **247,** 1797–1801 (2010).
32. Useinov, A., Deminov, R., Tagirov, L. & Pan, G. Giant magnetoresistance in nanoscale ferromagnetic heterocontacts. *J. Phys. Condens. Matter* **19,** 196215 (2007).
33. Tagirov, L. R., Vodopyanov, B. P. & Efetov, K. B. Ballistic versus diffusive magnetoresistance of a magnetic point contact. *Phys. Rev. B* **63,** 104468 (2001).
34. Useinov, A. N. *et al.* Mean-free path effects in magnetoresistance of ferromagnetic nanocontacts. *Eur. Phys. J. B* **60,** 187–192 (2007).
35. Useinov, N. Kh. Semiclassical Green's functions of magnetic point contacts. *Theor. Math. Phys.-Engl. Tr.* **183,** 705–714 (2015).
36. Useinov, A. *et al.* Impact of lattice strain on the tunnel magnetoresistance in Fe/insulator/Fe and Fe/insulator/La$_{0.67}$Sr$_{0.33}$MnO$_3$ magnetic tunnel junctions. *Phys. Rev. B* **88,** 060405(R) (2013).
37. Useinov, N. & Tagirov, L. Resonant magnetoresistance in asymmetrical double-barrier magnetic tunnel junction. *Physics Procedia* (Proceeding is accepted for publication, 20th International Conference on Magnetism, ICM-2015, Barcelona, 5–10 July, 2015).
38. Kubo, R., Kawabata, A. & Kobayashi, S. Electronic properties of small particles. *Ann. Rev. Mater. Sci.* **14,** 49–66 (1984).
39. Ghosh, S. K. Kubo gap as a factor governing the emergence of new physicochemical characteristics of the small metallic particulates. *Assam university journal of science and technology: Physical sciences and technology* **7,** 114–121 (2011).
40. Landau, L. D. & Lifshitz, E. M. *Quantum Mechanics*. Non-relativistic theory, 3d edn, Ch. 6, 133–159 (Pergamon press, 1991).
41. Yang, H. *et al.* Coexistence of the Kondo effect and a ferromagnetic phase in magnetic tunnel junctions. *Phys. Rev. B* **83,** 174437 (2011).
42. Tatara, G., Zhao, Y. -W., Muñoz, M. & García, N. Domain wall scattering explains 300 ballistic magnetoconductance of nanocontacts. *Phys. Rev. Lett.* **83,** 2030–2033 (1999).
43. Tagirov, L. R., Vodopyanov, B. P. & Garipov, B. M. Giant magnetoresistance in quantum magnetic contacts. *J. Magn. Magn. Mater.* **258,** 61–66 (2003).
44. Kugler, Z. *et al.* Temperature and bias voltage dependence of Co/Pd multilayer-based magnetic tunnel junctions with perpendicular magnetic anisotropy. *J. Mag. Mag. Mat.* **323,** 198–201 (2011).
45. Manchon, A. *et al.* Modelling spin transfer torque and magnetoresistance in magnetic multilayers. *J. Phys.: Condens. Matter* **19,** 165212 (2007).
46. Theodonis, I., Kalitsov, A. & Kioussis, N. Enhancing the spin-transfer torque through proximity of quantum well states. *Phys. Rev. B* **76,** 224406 (2007).
47. Chen, X., Zheng, Q.-R. & Su, G. Spin transfer and critical current for magnetization reversal in ferromagnet-ferromagnet-ferromagnet double-barrier tunnel junctions. *Phys. Rev. B* **78,** 104410 (2008).
48. Diao, Z. *et al.* Spin transfer switching in dual MgO magnetic tunnel junctions. *Apl. Phys. Lett.* **90,** 132508 (2007).
49. Theodonis, I. *et al.* Anomalous bias dependence of spin torque in magnetic tunnel junctions. *Phys. Rev. Lett.* **97,** 237205 (2006).
50. Kalitsov, A., Silvestre, W., Chshiev, M. & Velev, J. P. Spin torque in magnetic tunnel junctions with asymmetric barriers. *Phys. Rev. B* **88,** 104430 (2013).

## Acknowledgements

The work was supported by MOST (grants No. 104-2221-E-007-046-MY2, 103-2112-M-007-011-MY3) and RFBR (grant No. 14-02-00348-a). A. Useinov acknowledges partial support by the Program of Competitive Growth of Kazan Federal University.

## Author Contributions

C.H.L. proposed the conceptual idea of the work (highlight the role of nanoparticles into formation of the anomalous TMR and induced spin transfer torque); L.X.Y. and T.H.W. designed an experiments and fabricated the pMTJ samples with nanoparticles, they analyzed the data of magnetic properties, also proposed and discussed an idea of the electron transport model, which explain some experimental results but out of Coulomb blockade approach; A.U. and N.U. proposed the theoretical model and performed numerical simulations and its analysis. N.U. provided an exact analytical solution for double barrier system; A.U. and C.H.L. discussed and wrote the manuscript.

## Additional Information

**Supplementary information** accompanies this paper at http://www.nature.com/srep

**Competing financial interests:** The authors declare no competing financial interests.

**How to cite this article**: Useinov, A. *et al.* Anomalous Tunnel Magnetoresistance and Spin Transfer Torque in Magnetic Tunnel Junctions with Embedded Nanoparticles. *Sci. Rep.* **5,** 18026; doi: 10.1038/srep18026 (2015).

# Anomalous tunnel magnetoresistance and spin transfer torque in magnetic tunnel junctions with embedded nanoparticles: supplementary information

Arthur Useinov[1,2,3], Lin-Xiu Ye[2], Niazbeck Useinov[3], Te-Ho Wu[4], and Chih-Huang Lai[2]

## GENERAL MODEL OF THE MAGNETIC POINT CONTACT

One can imagine metallic point-contact removing some pieces of the barriers $L_{1,2} \to 0$, Fig. 1 or (Fig. S1**a**). The metallic nanoparticle (NP) becomes a part of the top (left) and bottom (right) metallic FM layers connecting them together (transmission $D_s \to 1$). The charge current of the metallic point-contact with orifice cross section can be calculated according Eq. S1, which was derived in assumption that electron energy is equal to $E_F$, while $E_F \gg k_B T$ [1]:

$$I_s = \frac{e^2 p_{F,s,\min}^2 a^2 V}{2\pi \hbar^3} \int_0^\infty dk \frac{J_1^2(ka)}{k} F_s(k, D_s, l_s),\qquad\text{(S1)}$$

$$F_s(k, D_s, l_s) = F_s^{bal}(D_s) + F_s^{df'}(k, D_s, l_s) + F_s^{df''}(k, D_s, l_s),\qquad\text{(S2)}$$

where $e$ and $l_s$ are electron charge and mean free path, respectively; $a$ is radius of the spot-like contact and $k_B$ is Boltzmann constant. $k$ is Fourier image of radial variable $\rho$, which is the coordinate of electron on the contact plane; $J_1(x)$ is Bessel function. $p_{F,s,\min} = \hbar k_{F,s,\min}$ is value of the Fermi momentum, where $k_{F,s,\min}$ is Fermi wavenumber which has to be minimal (min) value between $k_{F,s}^L$ and $k_{F,s}^R$; $F_s(k, D_s, l_s)$ contains integration over $\theta_{c,s}$ which is angle between *z*-axis and direction of the electron velocity, where index $c = L(R)$ determines the contact side. $D_s$ is transmission coefficient determined in range of standard definition of the quantum mechanics. Transmission usually is a function of the electron wavenumbers, $\theta_{c,s}$, applied voltage *V* and other possible variables. These variables have been defined from specification of the considered problems where point-contact area can be exchanged by quantum object, similar to the present problem.

All integrals inside Eq. S2 are written in relation to variable $\theta_{L,s}$ in order to simplify solution and integral limits. The total current is summation of the both spin components of the charge current $I = I_\uparrow + I_\downarrow$. The complete view of the first and second terms of Eq. S2 accessed in ref. 2 The first term has a simple form $F_s^{bal} = \langle \cos(\theta_{L,s}) D_s \rangle_{\theta_L}$ and is responsible for the ballistic and quantum-ballistic transports while $F_s^{df'}(k, D_s, l_s)$ and $F_s^{df''}(k, D_s, l_s)$ are responsible for quasi-ballistic ( $l_\uparrow > a, l_\downarrow < a$ and $l_\uparrow, l_\downarrow \approx a$ ) and diffusive regimes of

[1]Department of Physics, National Tsing Hua University, Hsinchu, Taiwan. [2]Department of Materials Science and Engineering, National Tsing Hua University, Hsinchu, Taiwan. [3]Institute of Physics, Kazan Federal University, Kazan, Russian Federation. [4]Graduate School of Materials Science, National Yunlin University of Science and Technology, Douliou, Taiwan. Correspondence and requests for materials should be addressed to C.-H.L. (email: chlai@mx.nthu.edu.tw)

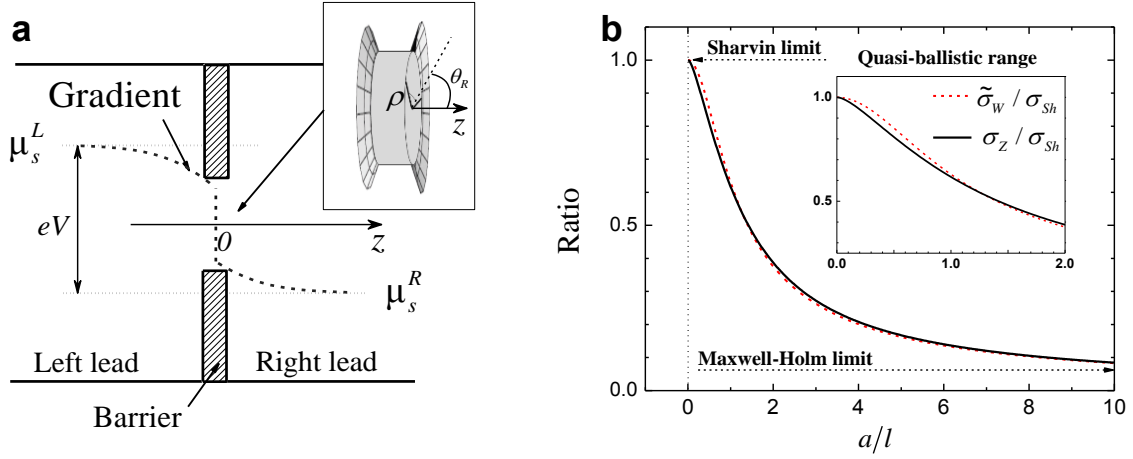

Figure S1. (**a**) Schematic plot of the point-like contact which is considered as an orifice in impenetrable wall ($a$-spot model); $\mu_s^{L(R)}$ is chemical potential. (**b**) shows numerical solution of the present model for $\sigma_Z/\sigma_{Sh}$ and refined Wexler's ratio $\tilde{\sigma}_W/\sigma_{Sh}$ as function of $a/l$.

---

the conduction (or Ohmic resistance, $l_\uparrow, l_\downarrow \ll a$). All three terms of Eq. S2 are the solutions of the system for the quasi-classical Green functions with quantum boundary conditions. The last term $F_s^{df''}(k, D_s, l_s)$ considers the gradient of the chemical potential nearby contact interface (for more technical details check ref. 1).

It is worth to note that $F_s^{bal}$ is not a function of $k$ and $l$, and this fact is an important fundamental quantum property of the electron in nanoscale. If take only $F_s(k, D_s, l_s) = F_s^{bal}(D_s)$ then Eq. S1 can be simplified since $\int_0^\infty J_1^2(x)/x\,dx = 1/2$, and:

$$\sigma_s = I_s/V = \frac{e^2}{h}\frac{k_{F,s,\min}^2 a^2}{2}\left\langle \cos(\theta_{L,s}) D_s \right\rangle_{\theta_L}. \tag{S3}$$

Equation S3 as a ballistic part of the general Eq. S1 shows validity of the Eq. 1. In case of symmetric *non*-magnetic point-like contact $k_F^L = k_F^R$, $\sigma_\uparrow = \sigma_\downarrow$, $D_\uparrow = D_\downarrow \to 1$, $F_s^{bal} = \left\langle \cos(\theta_{L,s}) \right\rangle_{\theta_L} = 1/2$, and finally: $\sigma_Z = \sigma_\uparrow + \sigma_\downarrow = (2e^2/h)(k_F^2 a^2/4)$ which coincides with Sharvin conduction limit[3,4] $\sigma_{Sh}$. Thus, Eq. S3 is some kind of extension of Sharvin conduction limit and characterizes the quantum conduction of the quantum system which is located inside the contact area. The quantum physics of the system can be accounted through analytical or numerical view of the transmission. The conductance (Eq. S3) is different from "classical ballistic" one for the general case where $D_s \neq 1$, the definitions in terms of "coherent", "direct" or "quantum-ballistic" limits are more appropriated. Furthermore, in order to analyze the solution in range of the complete expression (Eq. S1) in the limit of the *non*-magnetic symmetric point-contact, all terms in Eq. S2 were simplified:

$$\sigma_Z = 4\sigma_{Sh}\left(\frac{1}{4} - \int_0^\infty \frac{dx}{x}\frac{J_1^2(x)}{1+(xK)^2+\sqrt{1+(xK)^2}}\right) \tag{S4}$$

where $K = l/a$. In the limit when $K \to \infty$ ($a/l \to 0$), the integral is vanishing in Eq. S4 and conductance transformed into Sharvin limit.

Taking into account the exact integral's asymptotic for $K \to 0$ ($a/l \to \infty$):

$$\lim_{K \to 0} \int_0^\infty \frac{dx}{x} \frac{J_1^2(x)}{1+(xK)^2+\sqrt{1+(xK)^2}} = \frac{1}{4} - \frac{2}{3\pi}K$$

it is easy to obtain an exact diffusive solution (or Maxwell-Holm limit $\sigma_M$), $\sigma_Z \to \sigma_M = (8K/3\pi)\sigma_{Sh} = 2a/\rho_V$, where $\rho_V = \left(e^2 nl/\hbar k_F\right)^{-1} = \left(e^2 p_F^2 l/3\pi^2 \hbar^3\right)^{-1}$ is resistivity in volume [$\Omega \cdot$m] and $n = k_F^3/3\pi^2$ is electron density in metals. The dependence of the numerical ratio $\sigma_Z/\sigma_{Sh}$ on $a/l$ is very close to Wexler solution $\tilde{\sigma}_W/\sigma_{Sh} = \left(\frac{3\pi}{8K}\gamma(K)+1\right)^{-1}$ with Mikrajuddin's corrections[5,6], where $\gamma(K) \approx \frac{2}{\pi}\int_0^\infty e^{-K \cdot x} \text{sinc}(x) dx$. Figure S1**b** clearly reveals that our model shows good matching between Sharvin and Maxwell-Holm classical limits.

## TEMPERATURE DEPENDENCE

Considering the *T*-impact, it is noted that the ballistic conductance shows lack of dependence on *T*. As a result, the thermal heat occurs relatively far away from the contact area, i.e. on the distance larger than mean free path. However, Ohmic resistance (diffusive regime) depends on *T* since metal resistivity is sensitive to the temperature[4]. Yet the temperature impact for the simulating NP in range of double barrier system might be considered as indirect *T*-dependence of the transmission, especially for $k_n \approx 0.1-0.46 \text{Å}^{-1}$. Transmission $D_s$ is a function of $k_n$ and if the Fermi energy of NP is comparable with $k_B T$ then $k_n$ has the margin of the values, which significantly changes the conductance behavior due to conduction band broadening.

The key parameter of our model is $k_n$. The consideration of the system in terms of finite temperature depends on how corresponding energy of $k_n$ is compared to $k_B T$. In range of $k_n < 0.46 \text{Å}^{-1}$ (corresponding energy $E_n = \hbar^2 k_n^2/2m$) the thermal energy at room temperature is important, while for $0.1 < k_n < 0.26 \text{Å}^{-1}$ even a few Kelvins is important; therefore, we have to add additional integration over $k_n$ in the following form to get more correct conductance:

$$G_s \propto \int_{X_1}^{X_2} dx \int_0^{\theta_{\min}} \sin(\theta_s)\cos(\theta_s) D_s(\theta_s, k_{n,s}+x) d\theta_s , \qquad (S5)$$

where $X_{1(2)} = \mp\sqrt{2m k_B T/\hbar^2}$. With increasing temperature, the *T*-induced band broadening destroy TMR dips and peaks[7]. For example Fig. 3c, Fig. 3e and Fig. 3f in ref. 8 clearly show how the width of resonant peak increases with temperature at low voltages and its amplitude slowly decreases. However the resonant TMR peak has to be more stable for higher temperatures in contrast to TMR suppression since $k_n$ is larger (e.g. Fig. 3f in ref. 8 reveals this fact). Our simulations show classical dome-like TMR behaviors without anomalies when $k_n > 0.5 \text{Å}^{-1}$ and developed approach is valid at room temperature at this case.

## ACKNOWLEDGMENT

The work was supported by MOST, Taiwan (grants No.104-2221-E-007-046-MY2, 103-2112-M-007-011-MY3) and RFBR (grant No. 14-02-00348-a). A. Useinov acknowledges partial support by the Program of Competitive Growth of Kazan Federal University.

# REFERENCES

1. Useinov, N. Kh. Semiclassical Green's functions of magnetic point contacts. *Theor. Math. Phys.-Engl. Tr.* **183,** 705-714 (2015).

2. Useinov, A., Deminov, R., Tagirov, L. & Pan, G. Giant magnetoresistance in nanoscale ferromagnetic heterocontacts. *J. Phys. Condens. Matter* **19,** 196215 (2007).

3. Sharvin, Y. V. A possible method for studying Fermi surfaces. *Sov. Phys.* **21,** 655-656 (1965).

4. Timsit, R. S. Electrical conduction through small contact spots. *Electrical Contacts.* (Proceedings of the 50th IEEE Holm Conference on Electrical Contacts and the 22nd International Conference on Electrical Contacts). 184 -191 (2004).

5. Wexler, G. The size effect and the non-local Boltzmann transport equation in orifice and disk geometry. *Proc. Phys. Soc.* **89,** 927-941 (1966).

6. Mikrajuddin, A., Shi, F., Kim, H. & Okuyama, K. Size-dependent electrical constriction resistance for contacts of arbitrary size: from Sharvin to Holm limits. *Mater. Sci. Semicond. Process.* **2,** 321-327 (1999).

7. A.N. Useinov, and C.H. Lai, Tunnel magnetoresistance and temperature related effects in magnetic tunnel junctions with embedded nanoparticles. *SPIN* (accepted Nov. 2015).

8. Ciudad, D. *et al.* Competition between cotunneling, Kondo effect, and direct tunneling in discontinuous high-anisotropy magnetic tunnel junctions. *Phys. Rev. B* **85,** 214408 (2012).